\pdfoutput=1

\documentclass[12pt,preprint,superscriptaddress,aps,preprintnumbers,amsmath,amssymb,pral]{revtex4-1}
\usepackage{graphicx}
\usepackage{bm}
\usepackage{color}
\usepackage{amsmath,amssymb}	
\usepackage{hyperref}
\usepackage{setspace}
\usepackage{longtable}
\usepackage{booktabs}
\usepackage{comment}
\usepackage{array}
\usepackage{cancel}
\usepackage{enumitem}
\usepackage{slashed}
\usepackage{todonotes}

\def\slashchar#1{\setbox0=\hbox{$#1$} % set a box for #1
\dimen0=\wd0 % and get its size
\setbox1=\hbox{/} \dimen1=\wd1 % get size of /
\ifdim\dimen0>\dimen1 % #1 is bigger
\rlap{\hbox to \dimen0{\hfil/\hfil}} % so center / in box
#1 % and print #1
\else % / is bigger
\rlap{\hbox to \dimen1{\hfil$#1$\hfil}} % so center #1
/ % and print /
\fi}

\def\d{{\mit  \Delta} }

\def\beq{\begin{eqnarray}}
\def\eeq{\end{eqnarray}}

\begin{document}
%\newcolumntype{Y}{>{\centering\arraybackslash}p{23pt}} 

%%%%%%%%%%%%%%%%%%%%%%%%%%%%%%%%%%
%%%%%%%%%%% Title page %%%%%%%%%%%
%%%%%%%%%%%%%%%%%%%%%%%%%%%%%%%%%%

\preprint{IPMU21-0089}

\title{On the Detection of QCD Axion Dark Matter by Coherent Scattering
}

\author{Hajime Fukuda}
\affiliation{Theoretical Physics Group, Lawrence Berkeley National Laboratory, Berkeley,
 CA 94720, USA}
\affiliation{Berkeley Center for Theoretical Physics, Department of Physics,\\
 University of California, Berkeley, CA 94720, USA}

\author{Satoshi Shirai}
%\email[e-mail: ]{satoshi.shirai@ipmu.jp}
\affiliation{Kavli IPMU (WPI), UTIAS, The University of Tokyo, Kashiwa, Chiba 277-8583, Japan}

\date{\today}
\begin{abstract}
The QCD axion is a promising candidate of the dark matter.
In this paper, we discuss elastic scattering processes between nucleons and the QCD axion dark matter. We point out that the cross section can be enhanced by more than $\mathcal{O}(10^{25})$ by the coherent effect, compared to classical processes.
As a result, for example, we show that QCD axions may scatters at the sun with $\mathcal{O}(1)$ probability for $f_a \lesssim 10^{11}\,\text{GeV}$.
In addition, one may expect stimulated emission effects can also enhance the cross section because the number density of the axion DM is very large. The enhancement factor can be as large as another $\mathcal{O}(10^{25})$ and, if the factor exists, the force from the dark matter wind may be detected via e.g., torsion balance experiments. 
However, it is also found that there is a cancellation of the stimulated emission factor and the force is too small to be detected.
\end{abstract}

\maketitle

\section{Introduction}
\label{sec:introduction}

The absence of a Dark Matter (DM) candidate in the Standard Model (SM) invokes the strongest motivation for the extension of the SM.
The QCD axion \cite{Weinberg:1977ma,Wilczek:1977pj} is the leading candidate of the DM.
The QCD axion is a pseudo-Nambu-Goldstone Boson associated with the spontaneous breaking of Peccei-Quinn (PQ) symmetry.
Originally, it has been introduced as a solution to the strong CP problem; the unnaturally tiny value of the $\theta$ angle in the QCD, $|\theta| \lesssim 10^{-10}$\,\cite{Peccei:1977hh}.
Later, it has turned out that axions may be created in the early universe and be the DM in the present universe\,\cite{Preskill:1982cy,Abbott:1982af,Dine:1982ah}.

The axion is nothing but the dynamical QCD $\theta$ term.
The QCD axion $a$ inevitably couples to the gluon via the Chern-Simons term, $\frac{a}{f_a} G_{\mu\nu}^a \tilde{G}^{a\mu\nu}$, where $f_a$ is the axion decay constant.
The vacuum expectation value (VEV) of the axion field $\langle a \rangle$ determines to an effective $\theta$ angle as $\theta_\mathrm{eff} = \theta + \langle a \rangle/f_a$.
If the axion potential is solely generated by the QCD dynamics, 
the axion VEV exactly cancels the $\theta$ angle\,\cite{Witten:1980sp,Vafa:1984xg}.
In order for this axion solution to the strong CP problem to work well, the axion potential contributions other than QCD must be almost exactly zero.
This property of the axion is dubbed a ``shift symmetry''; the UV Lagrangian of the axion must be invariant under the constant shift of the axion field, $a \to a + a_0$, except for the axion-gluon interaction.
For the axion-gluon interaction, the shift is equivalent to the shift of the $\theta$ angle. Because the SM quarks have non-vanishing masses, this shift symmetry is explicitly broken, which introduces the axion potential.
In other words, the axion-gluon Chern-Simons term is the only term representing the breaking of the shift symmetry in the UV theory.

In the IR theory below the QCD scale, we can write down the effective interaction of the axion by using e.g., the chiral perturbation and the large $N_c$ expansion\,\cite{Witten:1980sp}.
We can estimate the axion mass, axion-hadron, and axion-photon couplings in the IR Lagrangian.
At the leading order of the chiral perturbation, the effective interactions of the axion respect the shift symmetry. 
For example, the axion-photon coupling, $a F_{\mu\nu} \tilde{F}^{\mu\nu}$, has the shift symmetry and
the axion-nucleon coupling is a derivative coupling, $\partial_\mu a \bar{N}\gamma^\mu\gamma_5N$, where $N$ is a nucleon Fermion and respects the shift symmetry. 

Many studies to date have aimed at detection based on this leading order interaction between the axion and Standard Model particles.
The QCD axion is strongly constrained from astrophysical observations such as stellar cooling\,\cite{Tanabashi:2018oca}. 
The current constraint indicates that  axion decay constant $f_a$ must be larger than $\sim 10^9\,\text{GeV}$.
In order to explore higher $f_a$ regions, experiments on the ground are necessary, and this research is being conducted vigorously.
In particular, many attempts have been made to observe the axions as the DM, as the axion is the most important and natural candidate of the DM.
See Refs.\,\cite{ADMX:2009iij,Budker:2013hfa,Caldwell:2016dcw,McAllister:2017lkb,HAYSTAC:2018rwy,CAPP:2020utb} for recent works. 
However, such experiments are not sensitive to most of the parameter regions predicted by the QCD axion DM.
It is thus important to investigate another detection method.

In this paper, we discuss the direct detection, {\it i.e.,} elastic scatterings, of the axion DM. Although the axion is very light and the recoil energy of each scattering is small, the number of the axion is large, and in principle, the direct detection of the axion can be important. However, in most of the previous studies, elastic scattering processes of axions have not been considered. This is because the cross section is considered to be small. The shift-symmetric interactions between axions and Fermions such as axion-nucleon interactions, $\partial_\mu a \bar{N}\gamma^\mu\gamma_5N$, are dependent on the spin of the Fermion. The scattering amplitudes are, when added up over multiple Fermion scatterers with unaligned spins, mostly canceled.

However, we point out that spin-independent interactions between axions and nucleons, $(a/f_a)^2 \bar{N}N$, actually do exist. Those interactions break the shift symmetry and are proportional to the shift-symmetry-violating parameter. It may be naively expected that
such interactions are suppressed by the axion mass in addition to the $f_a^{-2}$ suppression as
only the axion mass seems to violate the shift symmetry in the IR Lagrangian. 
However, this is not the case. In the leading order of the chiral perturbation and the large $N_c$ expansion, the shift-symmetry-violating parameters in the UV Lagrangian are the quark masses. Indeed, we find that the four-point interaction, $(a/f_a)^2 \bar{N}N$, is proportional to the quark mass and not very smaller than shift-symmetric interactions. 

Unlike spin-dependent amplitudes, spin-independent amplitudes are constructively summed over multiple Fermion scatterers within one Compton length of the axion. Cross sections, which are proportional to the amplitudes squared, are enhanced by the number of the scatterer compared to classical estimations. This is called the coherent enhancement effect and is included in the context of the direct detection of heavier mass dark matters\,\cite{Goodman:1984dc}. An important point is that the mass of the axion is much lighter and the Compton length of the DM axion is macroscopic. The resultant enhancement effect is expected to be gigantic.

Another important enhancement effect for the direct detection of the axion is the stimulated emission effect. 
As a rule of quantum mechanics, the scattering amplitude of the process $a+b \to c + d$, $\mathcal{M}(a b \to c d)$, is enhanced by the phase space number density of particles $d$ if $d$ is a Boson. For example,
\begin{align}
    \left|\mathcal{M}(a b \to c d)\right|^2 = \left|\mathcal{M}_0(a b \to c d)\right|^2 \times (1 + f_d),
\end{align}
where $\mathcal{M}_0$ is the amplitude without the background and $f_d$ is the phase space number density of the final state particle $d$.
Given that we know the DM energy density $\rho_\text{DM}$ and velocity $v_\text{DM}$, the phase space number density of the light DM such as an axion is very large. The phase space number density of the DM of a mass $m_\text{DM}$, $f_\text{DM}$, is
\begin{align}
    f_{\rm DM} \sim \frac{\rho_{\rm DM}}{m_{\rm DM}  \left(m_{\rm DM} v_\text{DM}\right)^3 } \sim 10^{27}\left(\frac{m_\text{DM}}{1~\mu{\rm eV}}\right)^{-4}.
\end{align}
One may expect that the cross section is enhanced by $f_\text{DM}$. If this is the case, it will enhance the rate of the direct detection of light Bosonic DMs such as axions very much. 
As we will see later, although the cross section is indeed enhanced, the recoil momentum is not enhanced by this stimulated effect.

The rest of this paper is organized as follows. First, in Sec\,\ref{sec:axionInteraction}, we review the axion interaction. We use the chiral perturbation to derive the effective Lagrangian for the axion and hadrons. In Sec.\,\ref{sec:axionDetection}, we discuss the detection method using the interaction. In quantum mechanics, forward scattering amplitudes, not amplitudes squared, may induce a physical effect. Because the axion DM is light and has wave nature, we discuss both forward scatterings and ordinary scatterings. In Sec.\,\ref{sec:conclusion}, we conclude the paper with discussing other possible applications. 

\section{Axion Interaction}
\label{sec:axionInteraction}

The Lagrangian of the QCD axion above the QCD scale is written as follows:
\begin{eqnarray}
\label{eq:axionLagrangian}
\mathcal{L} = \frac{1}{2}\partial_\mu a\partial^\mu a + \frac{\alpha_s}{8\pi} N_\text{DW} \frac{a}{f_a} G_{\mu\nu}^a \tilde{G}^{a\mu\nu} +
\frac{\alpha}{4\pi} c \frac{a}{f_a} F_{\mu\nu} \tilde{F}^{\mu\nu},
\end{eqnarray}
where $a, G$ and $F$ are the axion, gluon and photon field, respectively, and $N_\text{DW}$ and $c$ are $\mathcal{O}(1)$ integer and rational constant, respectively. For simplicity, we take $N_\text{DW} = 1$ and $c = 0$. Here, the origin of the axion field is chosen to absorb the $\theta$ angle in QCD. The axion may also couple to the SM Fermions $f$ via derivative couplings as
\begin{align}
    \frac{\partial_\mu a}{f_a}\bar{f}\gamma_5\gamma^\mu f.
\end{align}
This does not alter our discussion, because the derivative couplings respect the shift symmetry.

We now discuss the axion Lagrangian below the QCD scale.
The axion-gluon interaction term in the UV Lagrangian, Eq.\,\eqref{eq:axionLagrangian}, breaks the shift symmetry by the virtue of the non-zero masses of the SM quarks. As a result, in the IR Lagrangian, shift-symmetry breaking terms appear, being proportional to the quark masses. The typical example is the axion mass;
\begin{align}
\label{eq:axionMass}
    m_a^2 \simeq \frac{\Lambda^3}{f_a^2 }\frac{m_um_d}{m_u + m_d},
\end{align}
where $\Lambda$ is the QCD scale and $m_{u(d)}$ is the up (down) quark mass.

We point out that the effect of the shift symmetry breaking, {\it i.e.,} the quark mass in the UV Lagrangian, also appears in the axion-nucleon interaction in the IR Lagrangian. Using the chiral perturbation theory, we obtain a spin-independent interaction as
\begin{align}
\label{eq:spinIndependentInt}
    \Delta\mathcal{L} = \frac{a^2}{8f_a^2}\sum_{N=p,n}\delta m_N \bar{N}N,
\end{align}
where $\delta m_N$ is a parameter related to the quark mass contribution to the nucleon mass, $\delta m_N \sim \mathcal{O}(10)\,\text{MeV}$, and given in Eq.\,\eqref{eq:deltaM}.
Note that, as discussed in Appendix, the $\delta m_N \propto m_u m_d$ and represent the shift-symmetry breaking of the UV theory.
The detailed calculation is found in App.\,\ref{sec:appChiralPert}.

At the first glance, the same operator, $a^2 \bar{N}N$, comes from the axion-nucleon interaction usually considered,
\begin{align}
\label{eq:axionOrdinary}
    m_N \bar{N}\exp\left(2ia\gamma^5/f_a\right)N,
\end{align}
where $m_N$ is the mass of the nucleon. Indeed, if we expand this interaction in $a$, one finds the term $a^2 \bar{N}N$.
However, spin-independent interactions induced by Eq.\,\eqref{eq:axionOrdinary} is suppressed by the axion mass. 
This is because the interaction Eq.\,\eqref{eq:axionOrdinary} respects the shift symmetry. 
To see this point clearly, by the chiral rotations of nucleons, the interaction is rewritten as 
\begin{align}
    \frac{\partial^\mu a}{f_a}\bar{N}\gamma_\mu \gamma^5N,
\end{align}
which is explicitly shift-symmetry-preserving.
Although we may write the four-point correlation function for $aaNN$ with this term, it is obvious that the correlator is suppressed by the momentum of the axion. In terms of Eq.\,\eqref{eq:axionOrdinary}, this can be explained as follows; indeed the $a^2 \bar{N}N$ term exists in Eq.\,\eqref{eq:axionOrdinary}, but for the scattering ampitude for the soft axions, the diagram from $a^2 \bar{N}N$ cancels against the diagrams with two $a \bar{N}i\gamma^5N$ vertices which are also included in the expansion of Eq.\,\eqref{eq:axionOrdinary}. Because of the momentum suppression, in total, Eq.\,\eqref{eq:axionOrdinary} hardly contributes to the spin-independent cross section. On the other hand, as we discuss in App.\,\ref{sec:appChiralPert}, Eq.\,\eqref{eq:spinIndependentInt} does contribute to the $aaNN$ amplitude without the momentum suppression. 
This is due to the nature of Goldstone boson and the similar property of the pion is known as the soft-pion theorem\,\cite{Weinberg:1996kr}.

Let us discuss a non-relativistic scattering process between DM axions and the target nucleons, $a N \to a N$, by Eq.\,\eqref{eq:spinIndependentInt}.
Because Eq.\,\eqref{eq:spinIndependentInt} does not include $\gamma^5$ in the nucleon bilinears, the scattering amplitude is independent of the spin of the nucleon.
Thus, if multiple nucleons exist within one Compton length of the incoming axion, the phases of each scattering amplitude are the same. 
Because all the phases are the same, the total amplitude is added up constructively, being proportional to the number of nucleons. The cross section, which is proportional to the absolute value squared of the total amplitude, is then proportional to the number squared. The cross section is enhanced by the number of nucleons, compared to the classical scattering, where the total scattering cross section is the sum of each cross section. 
This is the coherent enhancement and common for low-energy scatterings in quantum mechanics. 
As the Compton length of the axion DM is macroscopic as
\begin{align}
    (m_a v_\text{DM})^{-1} \sim \frac{10^{-1}\,\text{eV}}{m_a}\,\text{cm},
\end{align}
the number of nucleons within the Compton length can be as large as $\gtrsim 10^{25}$, which leads to a significant enhancement of the cross section.

In order to quantitatively evaluate scattering amplitudes with the large coherent enhancement in the non-relativistic limit, it is convenient to use the first quantization formalism, as the process $a N \to a N$ is number-conserving. To derive a potential of the axion-nucleon system, we rewrite Eq.\,\eqref{eq:spinIndependentInt} using the number densities.
\begin{align}
\label{eq:SpinIndepIntNDensity}
    \mathcal{H}_I(x) = -\frac{n_a(x)}{4b m_\pi f_\pi f_a}\sum_{N=p,n}\delta m_N \left(n_N(x) + n_{\bar{N}}(x)\right),
\end{align}
where $\mathcal{H}_I(x) = -\Delta \mathcal{L}$ is the interaction Hamiltonian, $n_a$ is the number density of the axion, $n_{N}(n_{\bar{N}})$ is the number density of the (anti-)nucleon $N$ and $b \equiv \frac{2 m_a f_a}{m_\pi f_\pi}$, which reduces to $b \sim 1$ for the QCD axion. Therefore, the potential energy for each axion is
\begin{align}
\label{eq:axionPotential}
    V(x) = -\sum_{N=p,n}\frac{\delta m_N}{4b m_\pi f_\pi f_a} n_N(x),
\end{align}
where we assume $n_{\bar{N}} = 0$.
With this potential, we can write the Schr\"odinger equation for the axion.
\begin{align}
    \left(-\frac{\nabla^2}{2m_a} + V\right)\psi(x) = E \psi(x),
\end{align}
where $\psi(x)$ is the wavefunction of the axion and $E$ is the energy eigenvalue.
Solving the equation with appropriate boundary conditions, we can derive the scattering cross section for any nucleon configuration. 
In the Born approximation, the differential cross section is
\begin{align}
\label{eq:BornAppDiff}
    \frac{d\sigma}{d\Omega} \simeq \left(\frac{2m_a}{4\pi}\right)^2 \left|\int d^3 x e^{i\vec{q}\cdot\vec{x}} V(x)\right|^2,
\end{align}
where $\vec{q} = \vec{q}(\Omega)$ is the momentum transfer. For  the long-Compton-length limit, $q \lesssim m_a v_\text{DM} \to 0$, $\sigma \propto N_N^2$, where $N_N$ is the number of nucleons. 
This result reproduces the qualitative argument of the coherent enhancement above. 
We discuss possible ground experiments exploiting this potential for the axion DM detection in the next section.

To close this section, we make a few comments on other aspects of the interaction, Eq.\,\eqref{eq:SpinIndepIntNDensity}.
First, it can be interpreted as potential energies for nucleons. 
As in the case of the axion potential Eq.\,\eqref{eq:axionPotential}, the potential for the nucleon is
\begin{align}
    V_N(x) = -\frac{\delta m_N}{4 f_a^2} a^2(x) .
\end{align}
This induces the mass correction to the nucleons with an axion background.
In particular, this potential affects the mass difference between the neutron and proton and modifies the lifetime of the neutron\,\cite{Ubaldi:2008nf}.
For cosmological constraints, see Ref.\,\cite{Blum:2014vsa}.

Second, because the axion is a dynamical field, it also mediates the Yukawa potential between two nucleons via Eq.\,\eqref{eq:spinIndependentInt}, with an axion background $\langle a(x) \rangle = a_0$. The Lagrangian \eqref{eq:spinIndependentInt} is, with the background, read as
\begin{align}
    \Delta\mathcal{L} = \sum_{N=p,n}\frac{a_0 \delta m_N}{4f_a^2} a\bar{N}N.
\end{align}
This induces a long-range attractive force between the nucleons. 
The Yukawa potential for two $N$ with a  distance of $r$ is
\begin{align}
    V_Y(r) = - \left(\frac{\sqrt \pi r_N M_\text{Pl} a_0}{\sqrt 2 f_a^2}\right)^2 \frac{m_N^2}{8 \pi M_\text{Pl}^2}\frac{e^{-m_a r}}{r},
\end{align}
where $r_N \equiv \delta m_N / m_N$ and $M_\text{Pl}$ is the reduced Planck mass. 
This result is essentially the same as one given in Ref.\,\cite{Moody:1984ba}.

\section{Signal of Axion Scattering}
\label{sec:axionDetection}

In this section, we discuss experimental setups of the axion DM  direct detection to use the potential \eqref{eq:axionPotential}.
Because the Compton length of the axion DM is macroscopic and can be comparable to detector sizes, the wave nature of the axion is also important.
In addition to conventional ``hard" scattering, which is proportional to the squared of the scattering potential $|V|^2$, there is forward amplitude proportional to $V$, which corresponds to refraction of a wave and may provide observable signatures.
Here, we discuss both $\mathcal{O}(f_a^{-1})$ and $\mathcal{O}(f_a^{-2})$ effects, although it will turn out that the former effect is hardly possible to use with conventional detectors.

\subsection{$\mathcal{O}(f_a^{-1})$ effect}
Because the Compton length of the axion DM is macroscopic, the axion DM can be regarded as a wave. When a wave enters a medium, the wave is refracted. As we discuss in App.\,\ref{app:refraction},
the index of refraction $n$ is given as
\begin{align}
    n - 1 \simeq 1 \times 10^{-10}
    \left(\frac{\delta \rho}{0.01\,\text{g}/\text{cm}^3}\right) \left(\frac{10^{-3}}{v_\text{DM}}\right)^2 b^{-2}.
\end{align}
Here, 
\begin{align}
    \delta \rho = \sum_{N=p,n}\delta m_N n_N = \left(\frac{\delta m_p}{m_p} \cdot\frac{Z}{A} + \frac{\delta m_n}{m_n} \cdot\frac{A - Z}{A} \right)\rho,
\end{align}
where $Z(A)$ is the atomic (mass) number of the target medium and $\rho$ is the mass density of the medium.
The index is small, but it is not obvious that we may ignore it. 
Let us discuss how the index of refraction may contribute to the observational quantity in the following. 

Suppose an incoming axion DM is refracted on the surface of a target composed of nucleons. As the momentum of the axion is altered, a momentum of order $m_a v_\text{DM} (n - 1)$ is transferred to the target. 
Due to the motion of the solar system and the earth through the Milky Way ($v_{\oplus}  \sim 200$ km/s), there is a DM wind from the direction of travel of the solar system.
For the DM wind, one may expect that the net momentum transfer by the refraction is non-zero if we choose a target with an appropriate surface. 
Naively, for a target with surface area $S$ and mass $M$, the axion DM wind induces an acceleration $a$ on the target of order
\begin{align}
    a \stackrel{?}{\sim} \frac{\rho_\text{DM} v_\text{DM} |n - 1| S}{M} = 10^{-17} \left(\frac{|n - 1|}{10^{-10}}\right) \left(\frac{S}{\text{cm}^2}\right) \left(\frac{\text{g}}{M}\right)\,\text{cm}/\text{s}^2.
\end{align}
Although this acceleration looks small, it is large enough to be detected in an experiment like a torsion balance\,\cite{Hagmann:1998nz,*Hagmann:1999kf}. 
Thus, it is worth asking if the refraction indeed induces such large acceleration.

In fact, similar discussions have been made for the detection of the cosmic neutrino background (C$\nu$B)\,\cite{Fukugita:2003en}. 
The C$\nu$B is expected to be as cold as $\sim1.9\,\text{K}$ and thus its Compton length is also macroscopic. 
The weak Boson exchange processes induce a potential between neutrinos and matters.
If the neutrino is a Dirac fermion, the potential for a neutrino $\nu_i$ ($i = e, \mu$ and $\tau$) is given as\,\cite{Fukugita:2003en}
\begin{align} \label{eq:neutrino_potential}
    V_{\nu_i} = \frac{G_\mathrm{F} f_i}{\sqrt{2}} n_A.
\end{align}
Here, $G_\mathrm{F}$ is the Fermi constant, $n_A$ is the atomic number density in the target medium.
The parameter $f_i$ is a numerical constant given as
\begin{align}
    f_i = \begin{cases}
    Z - A & i \ne e\\
    3Z - A &  i = e.
    \end{cases}
\end{align}

As in the discussion of the axion DM, the neutrino potential contributes to the refractive index; the refractive index of the cosmic background neutrino in a medium is
\begin{align}
    \frac{n - 1}{f_i} \sim 10^{-5} \left(\frac{n_A}{6\times10^{23}/\text{cm}^3}\right) \left(\frac{0.01\,\text{eV}}{m_\nu}\right) \left(\frac{10^{-3}}{v_{\nu}}\right)^2.
\end{align}
This refraction index seems to be large.
Several proposals were, accordingly, made for the C$\nu$B detection\,\cite{Opher:1974drq,Stodolsky:1974aq,Lewis:1979mu}.

However, it has turned out that the refraction effect is hardly detectable\,\cite{Langacker:1982ih,Cabibbo:1982bb}, as the total momentum transfer is completely canceled for the refraction effect for spatially uniform C$\nu$B density.
The net momentum transfer from the neutrino background to a target per unit time, $\dot{\vec{P}}$, is given as
\begin{align}
    \dot{\vec{P}} & = - \dot{\vec{P}}_\nu = -i [H_I, \vec{P}_\nu] \nonumber \\
    &= -\int d^3 x V_\nu \vec{\nabla}n_\nu,
\end{align}
where $\dot{\vec{P}}_\nu$ is the momentum change of the neutrino background, $n_\nu$ is the number density of the C$\nu$B and $H_I$ is the interaction Hamiltonian given as $H_I = \int d^3 x V_\nu n_\nu$ and $V_\nu$ is the neutrino potential \eqref{eq:neutrino_potential}. 
This equation shows that as long as the neutrino background is homogeneous, $\nabla n_\nu = 0$, the total momentum transfer vanishes.

The same result holds for the axion DM. 
As long as the axion DM distribution is uniform, the static force by the refraction vanishes. 
One important difference is that axions are Bosons and can form lumps, where $\nabla n_a \ne 0$. 
For example, if axions move together as a wave packet with the width $\sim m_a v_\text{DM}$, the axion density may be non-homogeneous; $\nabla n_a \sim m_a v_\text{DM} n_a = v_\text{DM} \rho_\text{DM}$.
In this case, we estimate the refractive force as
%If this is the case, the momentum transfer reads
\begin{align}
    \dot{P} &\simeq - \int d^3 x V(x) v_\text{DM} \rho_\text{DM} \nonumber \\ &= \frac{v_\text{DM} \rho_\text{DM}}{8m_a f_a^2} M_r \nonumber \\
    &\simeq 2 \times 10^{-20}  \text{g}\cdot\text{cm}/\text{s}^{2} \times b \left(\frac{\rho_\text{DM}}{0.3\,\text{GeV}/\text{cm}^3}\right) \left(\frac{v_\text{DM}}{10^{-3}}\right) 
     \left(\frac{M_r}{0.01\,\text{g}}\right)
     \left(\frac{10^9\,\text{GeV}}{f_a}\right)
   ,
\end{align}
where $M_r \equiv \int d^3 x \delta \rho$ for the target. 
The resulting acceleration, $10^{-20}\,\text{cm}/\text{s}^{-2}$, is larger than the ultimate sensitivity of torsion balance experiments, $10^{-23}\,\text{cm}/\text{s}^{-2}$\,\cite{Hagmann:1998nz,*Hagmann:1999kf}. Nonetheless, conventional torsion balance experiments are not able to detect the signal\,\cite{Graham:2015ifn}, because the force depends on the dark matter local density configuration and has a strong time dependence with a frequency $\mathcal{O}(m_a v_\text{DM}) \sim  (m_a/1~\mu \text{eV})~ \text{MHz}$.
Moreover, as the axion is massive, the axion wave packet is dissipative and it is likely that $\nabla n_a \ll v_\text{DM} \rho_\text{DM}$ in reality. 
Therefore, we consider the axion DM detection with the refraction effect is hard. 

\subsection{$\mathcal{O}(f_a^{-2})$ effect}
In this subsection, we discuss elastic scattering processes by Eq.\,\eqref{eq:axionPotential}, which are suppressed by $f_a^{-2}$.
For quantum mechanical scatterings, we possibly have two enhancement effects, the coherent enhancement and the stimulated emission effect. As we have discussed in Sec.\,\ref{sec:axionInteraction}, the coherent effect is properly included if we solve the Schr\"odinger equation with the potential, Eq.\,\eqref{eq:axionPotential}. Because we cannot analytically solve the Schr\"odinger equation for generic potentials, we rely on the Born approximation instead, although the validity of the approximation must be discussed.
On the other hand, the stimulated emission effect is needed to be independently included in the estimation.

Roughly speaking, the stimulated emission effect comes from a matrix element $\mathcal{M}$ being evaluated as follows;
\begin{align}
\label{eq:stimulatedEmission}
    \mathcal{M} = \langle N_X - 1, N_Y + 1| \mathcal{S} | N_X, N_Y \rangle
    &\simeq \langle N_X - 1, N_Y + 1| -i\int dt H_\text{eff}| N_X, N_Y \rangle \nonumber \\
    & = \sqrt{N_Y + 1} \langle N_X - 1, 1| -i\int dt H_\text{eff}| N_X, 0 \rangle,
\end{align}
where $H_\text{eff}$ is the effective Hamiltonian, $\mathcal{S}$ is the S-matrix, $N_X$ is the number of initial particles in the initial state, and $N_Y$ is the number of the particle in the initial state with the same quantum number, {\it e.g.} a momentum, as the final particle. The matrix element is enhanced by a remainder of normalization factors, $\sqrt{N_Y + 1}$, as multi-particle states are normalized as
\begin{align}
    |N_Y\rangle = \frac{1}{\sqrt{N_Y!}} \left(a_Y^\dagger\right)^{N_Y} |0\rangle,
\end{align}
where $a_Y^\dagger$ is a creation operator. This enhancement is called the stimulated emission effect. For the scattering process, $N$ corresponds to the phase space number density $f(p)$, where we assume a spatially uniform distribution. Typically,
\begin{align}
    f(\vec{p}) \sim \frac{\rho_{\rm DM}}{m_a  \left(m_a v_\text{DM}\right)^3 } \sim 10^{11}\left(\frac{\rho_\text{DM}}{0.3\,\text{GeV}/\text{cm}^3}\right) \left(\frac{m_\text{DM}}{10^{-2}~{\rm eV}}\right)^{-4},
\end{align}
and we may naively expect a gigantic enhancement.

In the following, we include both effects to estimate the scattering effects of the axion DM. We first discuss the validity of the Born approximation and show when the approximation cannot be used. We then estimate the scattering cross section without the stimulated emission effect and see the difference when the stimulated emission effect is included. However, it will turn out that the stimulated emission effect makes no difference for detection experiments.
Hereafter, we assume that the scattering target is a sphere of a radius $R$ with constant nucleon densities $n_N$, unless otherwise noted.

First, we discuss the validity of the Born approximation; the Born approximation is not always valid even though the potential Eq.\,\eqref{eq:axionPotential} itself is much smaller than the energy scale of the scattering. This is because $R$ may be very large. The Born approximation is a good approximation as long as the scattering process is rare; {\it i.e.} the amplitude of the scattered wave is smaller than the incoming wave. For a very large $R$, this may not hold anymore, since any incoming wave eventually scatters if we take $R\to\infty$. 

Let us see what happens in the limit of $R \to \infty$. We may formally calculate the cross section according to Eq.\,\eqref{eq:BornAppDiff} for an arbitrary $R$.
First, let us estimate the cross section for $m_a v_\text{DM} R \gg 1$. The differential cross section
for $q \gg 1 / R$ is, in the limit of $R \to \infty$,
\begin{align}
    \frac{d\sigma}{d\Omega} \simeq \left(\frac{2m_a}{4\pi}\right)^2 \left|\frac{4\pi R\cos qR}{q^2} \cdot \frac{\delta\rho}{4bm_\pi f_\pi f_a}\right|^2.
\end{align}
The cross section is
\begin{align}
    \sigma &\sim \int_{-2(m_a v_\text{DM})^2}^{\frac1{R^2}} dq^2 \frac{d\sigma}{dq^2} 
     = R^2 \left|\frac{R}{v_\text{DM}} \frac{\delta\rho}{4bm_\pi f_\pi f_a} \right|^2.
\end{align}
On the other hand, for $m_a v_\text{DM} \ll 1 / R$, 
\begin{align}
    \sigma \simeq \pi R^2 \left(\frac{R^2\delta\rho}{6 f_a^2}\right)^2.
\end{align}
When we take $R \to \infty$, the cross section exceeds the geometrical cross section, $\sigma \sim \pi R^2$. It is expected that the Born approximation must be invalid as $R$ becomes bigger.

Indeed, as discussed in App.\,\ref{app:BornApp}, the condition for the Born approximation to be valid reads
\begin{align}
\label{eq:BornHighP}
    1 &\gg \frac{R}{v_\text{DM}}\frac{\delta \rho}{4b m_\pi f_\pi f_a} \nonumber \\
    & = \frac{4.3 \times 10^{-6}}{b} \left(\frac{R}{\text{km}}\right) \left(\frac{\delta\rho}{0.01\,\text{g}/\text{cm}^3}\right) \left(\frac{10^{-3}}{v_\text{DM}}\right) \left(\frac{10^9\,\text{GeV}}{f_a}\right),
\end{align}
for $m_a v_\text{DM} R \gg 1$ and 
\begin{align}
    1 &\gg \frac{R^2 \delta \rho}{8 f_a^2} \nonumber\\
    &= 1.4 \times 10^{-13}  \left(\frac{R}{\text{mm}}\right)^2 \left(\frac{\delta\rho}{0.01\,\text{g}/\text{cm}^3}\right)
    \left(\frac{10^9\,\text{GeV}}{f_a}\right)^2
\end{align}
for $m_a v_\text{DM} R \ll 1$.
As expected, the cross section is geometrical at the point where the inequality does not hold anymore.
For small size targets, the latter equation concludes that the Born approximation is always correct in realistic setups. However, for large-size targets, the former condition does not necessarily hold. In particular, if we consider scatterings between the axion DM and celestial bodies, the condition can be violated.

Suppose, for example, to take the sun as the target, $R_\odot \simeq 7 \times 10^5\,\text{km}$. The density of the sun varies from $\mathcal{O}(0.1)$ to $\mathcal{O}(100)\,\text{g}/\text{cm}^3$ for radii $R_\odot\mbox-0.1R_\odot$, respectively\,\cite{Basu:2009mi}. For this range of parameters, Eq.\,\eqref{eq:BornHighP} does not always hold for $f_a \lesssim 10^{11}\,\text{GeV}$; {\it i.e.} axions scatters at the sun with $\mathcal{O}(1)$ probability for $f_a \lesssim 10^{11}\,\text{GeV}$. For other DM candidates such as a DM with the weak interaction, for example, the expected scattering rate at the sun is much smaller. This scattering process at the sun is specific to the axion DM, where a large coherence enhancement effect exists.

As an application, we point out that {\it one may measure the axion-gluon coupling by electromagnetic-type axion detection experiments} such as Haloscope-type experiments, if the angular resolution is good enough. Haloscope-type experiments\,\cite{Sikivie:1983ip}, for example, detect axions by converting them to photons in a magnetic field. Thus, they are sensitive to the axion-photon coupling but not to the axion-gluon coupling. Let us assume that such an experiment has a good angular resolution and can point to the sun, whose angular diameter is $\delta \simeq 10^{-2}$ rad. Note that this, at least, requires to measure the momentum of the converted photon as accurately as $\mathcal{O}(\delta \cdot v_\text{DM}) \sim \mathcal{O}(10^{-5})$ and it will be challenging.

Due to the peculiar motion of the solar system, it is expected that there is a dark matter wind.
Because of this DM wind, the axion DM flux on earth is not isotropic.
If a sufficiently good angular resolution is available, it will be possible to verify this DM wind by observing these angular dependencies.
However, because axion DM with $f_a \lesssim 10^{11}\,\text{GeV}$ strongly interacts with the sun, much of the axion DM arriving from the sun's direction is particles that have been scattered by the sun.
Therefore, we expect the DM flux from the solar direction to be very different from the value predicted from the unscattered DM wind.
Furthermore, there should be annual modulation in these solar DM fluxes, which depends on the differential cross section of the sun and DM axion.
Therefore, in principle, it is possible to study the detailed interaction between the sun and axion. 
For example, if this scattering is isotropic, the DM flux should be constant regardless of the season.
In reality, we need to  solve the Schr\"odinger equation with an appropriate density profile of the sun to derive the angular dependence of the scattering of the axion at the sun.
A precise measurement of the angular dependence would allow us to probe the direct coupling between the axion and gluon, which plays an important role in establishing the ``QCD axion" as a solution of the strong CP problem.

A few remarks are made on other possible applications of geometrically large scatterings between axion DMs and celestial bodies.
First, as Eq.\,\eqref{eq:BornHighP} shows, for the given mass of an object, the denser the object is, the stronger the interaction is. 
Therefore, the DM axion strongly scatters with denser astrophysical objects such as white dwarfs and neutron stars even for much larger $f_a$.
Second, since the axion DM may interact with the celestial bodies with $\mathcal{O}(1)$ possibility, it may affect the motion of the celestial bodies. 
The result of a previous study\,\cite{Fukuda:2018omk} shows that the effect is too small to affect the motion of the solar system. 
For neutron star binary motions, gravitational dragging effects are dominant over the axion DM effect\,\cite{Pani:2015qhr}.

Let us now discuss a much smaller object, where the Born approximation is valid. 
Because the coherent effect is most significant for low energies, $\left|m_a v_\text{DM} \right| R \sim 1$, we assume $R = \left(m_a v_\text{DM}\right)^{-1}$.  The scattering cross section in the Born approximation is
\begin{align}
    \frac{d\sigma}{d\Omega} &= \eta^6 \left|\frac{\delta \rho}{3 b^2 m_\pi^2f_\pi^2 m_a v_\text{DM}^3}\right|^2 \nonumber \\
    &= 3.1 \times 10^{-20} \,\text{cm}^2 \times b^{-4} \eta^6 \left(\frac{\delta\rho}{0.01\,\text{g}/\text{cm}^3}\right)^2 \left(\frac{10^{-3}}{v_\text{DM}}\right)^6 \left(\frac{10^{-2}\,\text{eV}}{m_a}\right)^2,
\end{align}
where $\eta = m_a v_\text{DM} R$.
 The scattering cross section is not very small, but each momentum transfer is as small as of order $\mathcal{O}(m_a v_\text{DM})$. We need to detect the collective force on the target, such as the acceleration on the target\,\cite{Graham:2015ifn}. 
 Let $f(\vec{p})$ denote the phase space density of the DM axion of a momentum $\vec{p}$, where we assume the DM is spatially uniform. 
 Without the stimulated emission effect, the force on the target $\vec{F}$ is written as
\begin{align}
\label{eq:forceClassical}
    \vec{F} &
    = \int d^3 p_i d^3p_{f} \delta(|\vec{p_f}| - |\vec{p_i}|) \frac{|\vec{p_i}|}{m_a} (\vec{p_i} - \vec{p_f}) \frac{d\sigma}{d\Omega_{if}} f(\vec{p_i}),
\end{align}
where $\Omega_{if}$ is the solid angle between $\vec{p}_i$ and $\vec{p}_f$.
In this case, the magnitude of the force induced by the DM wind is estimated as
\begin{align}
    F &\simeq 4\pi \frac{d\sigma}{d\Omega} \rho_\text{DM} v_\text{DM}^2\nonumber \\
    &\simeq 1.9 \times 10^{-33}\,\text{N}\,\times b^{-4}\eta^6\left(\frac{\delta\rho}{0.01\,\text{g}/\text{cc}}\right)^2 \left(\frac{\rho_\text{DM}}{0.3\,\text{GeV}/\text{cc}}\right) %\times \nonumber\\ &
    \left(\frac{10^{-3}}{v_\text{DM}}\right)^4 \left(\frac{10^{-2}\,\text{eV}}{m_a}\right)^2.
\end{align}
The acceleration induced by this force is
\begin{align}
\label{eq:classicalAcc}
    a \simeq& 5.8 \times 10^{-30} \,\text{cm}/\text{s}^2 \times b^{-4} \eta^3\left(\frac{\delta\rho}{0.01\,\text{g}/\text{cm}^3}\right)^2 \left(\frac{\rho_\text{DM}}{0.3\,\text{GeV}/\text{cm}^3}\right) \times \nonumber\\&
    \left(\frac{\rho}{1\,\text{g}/\text{cc}}\right)^{-1} \left(\frac{10^{-3}}{v_\text{DM}}\right) \left(\frac{m_a}{10^{-2}\,\text{eV}}\right),
\end{align}
where $\rho$ is the mass density of the target.
Given that the ultimate sensitivity of the torsion balance experiment is of order $10^{-23}\,\text{cm}/\text{s}^2$, the expected acceleration by the axion wind is too small.

As we have discussed, in quantum mechanics, the stimulated emission effect must be included. 
It is worth asking if the stimulated emission effect may enhance the acceleration. Again assuming the spatial distribution of the axion is uniform, the cross section is enhanced with the stimulated emission effect, 
\begin{align}
    \frac{d\sigma}{d\Omega_{if}} \to \frac{d\sigma}{d\Omega_{if}} (1 + f(\vec{p_f})).
\end{align}
Typically, we have the enhancement factor as:
\begin{align}
    f(\vec{p}) \sim f_0 = \frac{\rho_{\rm DM}}{m_a  \left(m_a v_\text{DM}\right)^3 } \sim 10^{11}\left(\frac{\rho_\text{DM}}{0.3\,\text{GeV}/\text{cm}^3}\right) \left(\frac{m_\text{DM}}{10^{-2}~{\rm eV}}\right)^{-4},
\end{align}
Therefore we may naively expect an acceleration of order
\begin{align}
    a_\text{QM} \stackrel{?}{=} f_0 a \sim 10^{-19}\,\text{cm}/\text{s}^2,
\end{align}
which is large enough to be detected.

However, it turns out that the enhancement effect actually vanish for DM detection.
With the stimulated emission effect included, Eq.\,\eqref{eq:forceClassical} should be written as
\begin{align}
\label{eq:forceQuantum}
    \vec{F}_\text{QM} &
    = \int d^3 p_i d^3p_{f} \delta(|\vec{p_f}| - |\vec{p_i}|) \frac{|\vec{p_i}|}{m_a} (\vec{p_i} - \vec{p_f}) \frac{d\sigma}{d\Omega_{if}} f(\vec{p_i}) (1 + f(\vec{p_f})).
\end{align}
The difference between  Eq.\,\eqref{eq:forceClassical} and Eq.\,\eqref{eq:forceQuantum} is
\begin{align}
\label{eq:forceDiff}
    \delta\vec{F} &
    = \int d^3 p_i d^3p_{f} \delta(|\vec{p_f}| - |\vec{p_i}|) \frac{|\vec{p_i}|}{m_a} (\vec{p_i} - \vec{p_f}) \frac{d\sigma}{d\Omega_{if}} f(\vec{p_i}) f(\vec{p_f}).
\end{align}
However, we observe this integrand is odd under exchange $\vec{p}_i \leftrightarrow \vec{p}_f$ and $\delta \vec{F} = 0$ after the integration.
Therefore there is no net effect from the stimulating emission.

Since we conclude a large cancellation of a factor $f_0 \gg \mathcal{O}(10^{10})$, we need to be careful about whether there is any loophole in this argument or any factor left uncancelled. First, we stress that the stimulated emission effect of $\sqrt{N_Y + 1}$ enhancement is valid only in perturbation theory.
As we have shown in Eq.\,\eqref{eq:stimulatedEmission}, the stimulated emission relies on the perturbation; the enhancement factor, $\sqrt{N_Y + 1}$, comes from the perturbative expansion of the S-matrix. 
For generic cases including Eq.\,\eqref{eq:forceQuantum}, we may need to use a density matrix as the DM axion is not in a pure state, but the origin of the enhancement factor $f(\vec{p}_f)$ is the same; it is from the perturbative expansion of the S-matrix. 
Thus, once the enhancement effect becomes so large that the scattering process would be larger than the unitarity bound, a naive use of the stimulated emission effect fails. 
The physical interpretation of this is that multiple scatterings such as $N_X, N_Y \to N_X - n_m, N_Y + n_m$ for $n_m \gg 1$ occur.
For the case in our interest, the rate is much smaller than the unitarity bound and our present treatment of the stimulated effect is still valid. 
Second possible loophole is that we have assumed that the axion distribution is spatially uniform to show the cancellation. In our analysis, we have assumed $R \lesssim \left(m_a v_\text{DM}\right)^{-1}$ and the assumption may be justified.

\section{Conclusion and Discussion}
\label{sec:conclusion}
In this paper, we have discussed the spin-independent interaction of the QCD axion. Contrary to a naive expectation, the size of the interaction is not suppressed more than $\mathcal{O}(f_a^{-2})$. In particular, we focus on the axion DM and estimate the axion-nucleon scattering effect.

The forward scattering process by the spin-independent interaction induces the index of refraction for DM axions. At first glance, it causes acceleration on targets, since momenta are transferred for each refraction process. However, the acceleration from the refractive effect is cancelled after integrating over all the surface of the target, if the axion DM density is spatially uniform.

As the DM axion has a macroscopic Compton length, the coherent enhancement effect on the scattering is very large. On one hand, the scattering cross section between gigantic objects as celestial bodies, including the sun, can be geometrically large. The scattering may be a good tool to measure axion-gluon coupling.

On the other hand, the scattering cross section between smaller objects such as a test target on the ground can be enhanced by the coherent effect.  
However the acceleration to the target by the axion DM wind is smaller than the ultimate sensitivity of the torsion balance experiment. 
Although the quantum stimulated emission effect can enhance the scattering cross section, we have found that the stimulated emission effect is canceled after integrating over the phase space of the axion DM.

We have left a few points as open questions. 
First, it is not clear that whether any other experimental setups can detect the acceleration better. 
In addition to the static acceleration by elastic scatterings, time-varying acceleration due to the refractive effect with inhomogeneous DM density can exist.
We need a new type of experiment to detect such signals. 
Second, we have not discussed astrophysical signals by the spin-independent interaction. 
In particular, we have pointed out that interactions with denser objects such as white dwarfs or neutron stars are very strong. 
It is worth asking if the impact between these objects and the axion DM, which may form denser clusters, may cause any signal such as seismic waves.

Finally, let us comment on the case of the more generic axion-like particle.
Although our present analysis focuses on the QCD axion, its application to axion-like particle case that the relation between the gluon coupling and mass is not correlated, is straightforward.
One can simply change the value of $b \equiv \frac{2 m_a f_a}{m_\pi f_\pi}$ in the above expressions.
A model with $b \ll 1$ for small $f_a$ may be already tested by the torsion balance experiments.

%%%%%%%%%%%%%%%%%%%%%%%%%%%%%%%%%%%%%%
\vspace{-.4cm}  
\begin{acknowledgments}
\vspace{-.3cm}
HF would like to thank Markus Luty for useful discussion.
This work is supported by Grant-in-Aid for Scientific Research from the Ministry of Education, Culture, Sports, Science, and Technology (MEXT), Japan, 17H02878, 18K13535,  20H01895, 20H05860 and 21H00067 (S.S.), by World Premier International Research Center Initiative (WPI), MEXT, Japan (S.S.), and by the Director, Office of Science, Office of
High Energy Physics of the U.S. Department of Energy under the
Contract No. DE-AC02-05CH11231 (H.F.).
\end{acknowledgments}
%%%%%%%%%%%%%%%%%%%%%%%%%%%%%%%%%%%%%%

%%%%%%%%%%%%%%%
\appendix
%%%%%%%%%%%%%%%
\section{Chiral Perturbation for $\text{U}(n)_L\times\text{U}(n)_R \to \text{U}(n)_V$}
\label{sec:appChiralPert}
In this section, we derive the chiral effective Lagrangian for two-flavor QCD, $\text{U}(n)_L\times\text{U}(n)_R \to \text{U}(n)_V$ for $n = 2$, based on Ref.\,\cite{Weinberg:1996kr}. We assume the large $N_c$ limit\,\cite{Witten:1980sp} as well. Note that we adopt a notation that the pion decay constant $f_\pi \simeq 93\,$MeV, which differs from the notation in Ref.\,\cite{Weinberg:1996kr} by factor $2$.

\subsection{Goldstone degrees of freedom}
In the effective Lagrangian, we may regard the coset space of the symmetry, $G/H$, where $G = \text{U}(2)_L\times\text{U}(2)_R$ and $H = \text{U}(2)_V$, as the Goldstone Boson degrees of freedom. In order to write down the Lagrangian invariant under the group $G$, we need to know the transformation rule for the Goldstone Bosons under $G$. We start to write the transformation rule for the quark field:
\begin{align}
\label{eq:linearRealization}
    q\equiv \begin{pmatrix}u\\d\end{pmatrix} \to \exp\left[i\sum_{a} \left(\theta_a^V \lambda_a+ \theta_a^A\lambda_a\gamma_5\right) \right] q,
\end{align}
where $\lambda_a$ is the $\text{U}(2)$ generators normalized as $\text{Tr}(\lambda^a\lambda^b) = \delta^{ab}$ and $\theta^V$ and $\theta^A$ are the parameters of $G$. Because $\theta^V$ corresponds to $H$ and $\theta^A$ does to the coset space, we may map the Goldstone Bosons $\xi_a(x)$ to $\theta^A_a$. For convenience, we define $\gamma(\xi)$, an element of $G$:
\begin{align}
\gamma(\xi(x)) \equiv \exp\left[-i \gamma_5 \sum_a \xi_a(x)\lambda_a\right].
\end{align}
For any constant $g\in G$, $g\gamma(\xi)$ is the element of $G$. We can define $\xi'$ and $h(g,\xi)$ such that $g\gamma(\xi)$ corresponds to an element in the right coset space, $\gamma(\xi')$, up to the right action of some element in $H$, which in general depends on $\xi$ and written as $h(g,\xi)$:
\begin{align}
g\gamma(\xi(x)) = \gamma(\xi'(x)) h(g, \xi(x)).
\end{align}
We may impose on $\xi'$ a condition that $\forall h_0 \in H$, $h(h_0, \xi(x)) = h_0 = \text{const.}$ because the symmetry $H$ should be linearly realized. We then assume the transformation rule for $\xi$ under $G$ is given as this $\xi'$; namely,
\begin{align}
\gamma(\xi) \to \gamma(\xi') = g\gamma(\xi)h(g, \xi)^{-1}.
\end{align}
This is the non-linear realization of $G$ on the Goldstone Bosons $\xi$.

Using $\gamma$, we may construct a non-linear realization base for the quark. From $q$, which transforms as Eq.\,\eqref{eq:linearRealization} and called the linear base, we define a new field $\tilde{q}$:
\begin{align}
q\equiv \gamma \tilde{q}.
\end{align}
This is the non-linear realization base for $q$.
Under the action of $G$,
\begin{align}
    \tilde{q} \to \tilde{q}' = h(g, \xi) \gamma^{-1} g^{-1} g q = h(g, \xi)\tilde{q}.
\end{align}
Indeed, $G$ non-linearly acts on $\tilde{q}$.
Since the action of $G$ on $\tilde{q}$ is always vectorial, we may regard $\tilde{q}$ does not contain the Goldstone degree of freedom. In particular, the quark condensation should be written as
\begin{align}
\label{eq:quarkCondensation}
\bar{\tilde{q}}_i \tilde{q}_j = -\Lambda^3\delta_{ij},
\end{align}
after integrating out UV degrees of freedom. Here, we call $\Lambda$ the QCD scale.

We may separate the left and right handed part of $\gamma(\xi)$ as $\gamma(\xi) = \beta(\xi)P_L + \beta(\xi)^{-1}P_R$, where $\beta(\xi) \equiv \exp(-i \sum_a \xi_a\lambda_a)$. The transformation rule for $\beta(\xi)$ is written as 
\begin{align}
L \beta(\xi) &= \beta(\xi') h(\xi, g)
\end{align}
or
\begin{align}
R \beta(\xi)^{-1} &= \beta(\xi')^{-1} h(\xi, g),
\end{align}
where
\begin{align}
    L &= \exp\left(i\sum_a \theta_a^{L}\lambda_a\right),\\
    R &= \exp\left(i\sum_a \theta_a^{R}\lambda_a\right).
\end{align}
Here, $g = \exp\left[i\sum_{a} \left(\theta_a^V \lambda_a+ \theta_a^A\lambda_a\gamma_5\right) \right]$ and $\theta^{L(R)} = \theta^V\pm\theta^A$. Using $\beta(\xi)$, we can define a new field $U(\xi)$;
\begin{align}
    U(\xi)\equiv\beta(\xi)^2 = \exp(-2i \sum_a \xi_a\lambda_a).
\end{align}
The transformation rule of $U(\xi)$ is now simple;
\begin{align}
    U(\xi) \to U(\xi') \equiv L U(\xi) R^{-1},
\end{align}
Note that the action of $G$ onto $U$ is non-linear as well, but it is constant in the spacetime.

\subsection{Lagrangian for the Goldstone Bosons}
We use $U$ to construct a $G$-invariant Lagrangian for the Goldstone Bosons. With the breaking of $\text{U}(1)_A$ in the quantum level, the most generic Lagrangian is, in the leading order of the large $N_c$ expansion and the chiral perturbation\,\cite{Weinberg:1996kr,Witten:1980sp},
\begin{align}
    \mathcal L = -\frac{f_\pi^2}{4} \text{Tr}\left(\partial_\mu U\partial^\mu U^\dagger\right) - \frac{a_A}{N_c}\left(-i\ln\det U - \theta\right)^2,
\end{align}
where $f_\pi$ is the pion decay constant, $a_A$ is a constant parameter of mass dimension $4$ with and $\theta$ is the QCD vacuum angle. We may assume $\theta = 0$, but keep it here for the discussion of the axion field later; in the presence of the axion with the decay constant $f_a$ and the domain wall number $N_{DW}$, we just replace $\theta\to \theta + N_{DW}\frac{a}{f_a}$.

Let us write down the canonical form of the Lagrangian for the Goldstone Bosons. Expanding the Lagrangian up to the second order of $\xi^a$, we obtain
\begin{align}
    \mathcal{L} = -f_\pi^2(\partial\xi)^2 - \frac{a_A}{N_c}\left(-2\sqrt{2}\xi^0 - \theta\right)^2.
\end{align}
We can use a base 
\begin{align}
(\eta', \pi^i) = \sqrt{2}f_\pi(\xi^0, \xi^i)    
\end{align}
for $i = 1, 2, 3$ as a canonically normalized Goldstone Bosons. Note that $\pi^3$ is usually called $\pi^0$.

We may include the explicit symmetry breaking term, the quark mass. In the UV Lagrangian, the quark mass term is
\begin{align}
\mathcal{L}_{M, \text{UV}} = -\bar{q} M_q q,
\end{align}
where $M_q \equiv \text{diag}(m_u, m_d)$. For the IR Lagrangian, we integrate out quarks by using the quark condensation, Eq.\,\eqref{eq:quarkCondensation}.
Replacing $\bar{\tilde{q}}_i\tilde{q}_j \to -\Lambda^3 \delta_{ij}$ and $\bar{\tilde{q}}_i\gamma_5\tilde{q}_j \to 0$, we obtain
\begin{align}\label{eq:meson_mass}
    \mathcal{L}_{M, \text{IR}} = \Lambda^3 \text{Tr}(\gamma(\xi)M_q\gamma(\xi)) = \frac{\Lambda^3}{2}\text{Tr}\left[(U + U^\dagger)M_q\right]
\end{align}
This term gives the mass to the (pseudo) Goldstone Bosons. Using the canonical basis, we obtain that the pion mass is
\begin{align}
    m_\pi^2 = \frac{(m_u + m_d)\Lambda^3}{f_\pi^2},
\end{align}
for instance.

\subsection{Interaction between the nucleon and the Goldstone Bosons}
Using the chiral perturbation, we can also write the interaction terms between the nucleon and the Goldstone Bosons. We know the nucleons, $\tilde{N} = (p, n)$ transforms as doublet under $H$. Thus, we may assume that $\tilde{N}$ is the non-linear realization base of the nucleon, which transforms as $\tilde{N} \to \tilde{N}' = h(\xi, g)\tilde{N}$ under $\forall g \in G$. The linear base can be defined as
\begin{align}
    N \equiv \gamma(\xi) \tilde{N}
\end{align}
so that the action of $\text{SU}(2)_{L(R)}$ on $N$ is the $\text{SU}(2)$ rotation for the left(right)-handed Weyl component of $N$. Using the linear and non-linear basis, we may write the most generic $G$-invariant Lagrangian for the nucleon and the Goldstone Bosons in the leading order of the chiral perturbation as
\begin{align}
    \mathcal{L} = \bar{N}i\slashed{\partial}N - m_N \bar{\tilde{N}}\tilde{N} + \frac{1}{2}(1 - g_A) 
    i\bar{N}\gamma^\mu\left(U\partial_\mu U^{-1} P_L + U^{-1}\partial_\mu U P_R\right)N,
\end{align}
where $m_N$ is the nucleon mass in the absence of the explicit $G$ breaking, and $g_A$ is the axial current coupling constant.
The physics is independent of whether we use $N$ or $\tilde{N}$, but we need to use only either one of them. We here rewrite the Lagrangian in terms of $\tilde{N}$.
\begin{align}
    \mathcal{L} = \bar{\tilde{N}}i\slashed{\partial}\tilde{N} - m_N \bar{\tilde{N}}\tilde{N} +  
    i\bar{\tilde{N}}\gamma^\mu\left(\frac{1 + g_A}{2}\gamma(\xi)^{-1}\partial_\mu\gamma(\xi) + \frac{1 - g_A}{2}\gamma(\xi)\partial_\mu\gamma(\xi)^{-1}\right)\tilde{N}.
\end{align}
Note that $g_A$ satisfies the so-called Goldberger-Treiman relation\,\cite{Goldberger:1958vp}.

This Lagrangian is $G$-invariant so that the Goldstone Boson couplings are all derivative. However, as we have discussed, $G$ is explicitly broken by the quark mass term and the anomaly. The latter equally contributes to the axion and the Goldstone Boson and we ignore it in this paper. 
The former effect can be included in the effective Lagrangian by rewriting the quark mass terms in terms of nucleon fields.
Let us define the nucleon form factor $f_{Tq}^N$ as
\begin{align}
    m_N f_{Tq}^N = \langle N | m_q \bar{\tilde{q}} \tilde{q} | N \rangle
\end{align}
for $q = u, d$. Let us focus on the neutral component of mesons in the following; $\xi^{1, 2} = 0$. We may rewrite the quark mass in terms of the IR degrees of freedom as follows;
\begin{align}
    m_q \bar{q} q \to m_N f_{Tq}^N \bar{\tilde{N}} \gamma^2(\xi)_{qq} \tilde{N}.
\end{align}
In particular, we are interested in spin-independent interactions, terms without $\gamma_5$ in nucleon bilinears. For these terms, the quark mass correction for the nucleon interaction can be written as
\begin{align}
\label{eq:non_deriv_meson}
    \delta\mathcal{L} = -\sum_{N = p, n} \frac{m_N}{2} \text{Tr}\left[F^N (U + U^\dagger)\right]\bar{\tilde{N}}\tilde{N},
\end{align}
where
\begin{align}
    F^N = \text{diag}(f_{Tu}^N, f_{Td}^N).
\end{align}
For the vacuum, $U \to 1$, this terms play a role of the mass correction to nucleons by quark masses. In particular, we define
\begin{align}
    \delta m_N \equiv m_N \text{Tr}(F^N). 
\end{align}
$f_{Tq}^N$ is estimated by lattice simulations\,\cite{Abdel-Rehim:2016won} and $\delta m_N \sim 40\,\text{MeV}$.

\subsection{Inclusion of the Axion}

Finally, we include the axion in the chiral effective theory. For simplicity, we assume the KSVZ-type axion\,\cite{Kim:1979if,*Shifman:1979if}, where the axion does not have the additional derivative coupling between the quarks. The kinetic mixing between mesons and an axion is therefore vanishing. The axion enters the effective Lagrangian as replacing $\theta$ by $\theta - a / f_a$, where $f_a$ is the axion decay constant and the domain wall number is assumed to be unity.

To simplify the analysis, we assume
the $1/N$ contribution for the $\eta'$ mass is much larger than the quark mass contribution, which is phenomenologicaly sound, $m_{\eta'}^2 \gg m_\pi^2$\,\cite{Tanabashi:2018oca}, and assumed by many references like Ref.\,\cite{Weinberg:1996kr}. With this assumption, we can just replace $\xi^0 \to \frac{a}{2\sqrt{2}f_a}$ to integrate out $\eta'$.

In this paper, we focus on the axion and nucleon as the degrees of freedom; in particular, $aaNN$ coupling is of phenomenological importance. If we assume $m_u = m_d$, it is obtained from Eq.\,\eqref{eq:non_deriv_meson} by taking the second order of $a$ as
\begin{align}\label{eq:non_deriv_axion}
    \delta \mathcal{L} = \frac{a^2}{8f_a^2}\left(\delta m_p\bar{p}p +\delta m_n\bar{n}n\right).
\end{align}
This equation agrees with the result in Ref.\,\cite{Hook:2017psm}.

For generic quark masses, results are following; the axion mass is
\begin{align}
    m_a^2 = \frac{\Lambda^3}{8 f_a^2 f_\pi^2}\left(f_+ - g\right),
\end{align}
and the interaction is Rq.\,\eqref{eq:non_deriv_axion} with
\begin{align}
    \delta m_N &\to \left(j_- f_{Tu}^N + j_+ f_{Td}^N\right)m_N,
\end{align}
where
\begin{align}
    f_\pm &\equiv (4f_a^2 \pm f_\pi^2) (m_u + m_d),\\
    g &\equiv \sqrt{-64 f_a^2f_\pi^2 m_um_d +f_+^2}, \\
    h &\equiv \sqrt{16f_a^2f_\pi^2(m_u - m_d)^2 + (f_- + g)^2} ,\\
    j_\pm &\equiv \left(\frac{f_- + g \pm 8f_a^2 (m_u - m_d)}{h}\right)^2.
\end{align}
One can confirm that the axion-proton coupling is vanishing in the limit of $m_u \to 0$, as $f_{Tu}^N \to 0$.
Note that for $f_a \gg f_\pi$, the above equations can be reduced to
\begin{align}
    m_a^2 &\simeq \frac{\Lambda^3}{f_a^2 }\frac{m_um_d}{m_u + m_d} ,\\
    \label{eq:deltaM}
    \delta m_N &\to \left[\left(\frac{2 m_d}{m_u + m_d}\right)^2 f_{Tu}^N +
   \left(\frac{2 m_u}{m_u + m_d}\right)^2 f_{Td}^N\right]m_N.
\end{align}
These interactions for the axion have been not explicitly explored before, but are implicitly used in contexts such as Ref.\,\cite{Moody:1984ba,Ubaldi:2008nf,Hook:2017psm}. For an application for the axion in this context, see Ref.\,\cite{Blum:2014vsa}.

\section{Refractive Index}
\label{app:refraction}
In this section, we discuss the index of refraction for the axion DM.
The index of refraction of a medium is defined as the ratio of momenta in a vacuum and in the medium.
Suppose an axion, whose momentum in the vacuum is $p^\mu = (E, \vec{k})$ so that $E^2 = k^2 + m_a^2$, goes into a medium and the four momentum is modified as $p^\mu = (E, \vec{k}')$. The index of refraction $n$ is then given as $\vec{k}' = n \vec{k}$.

In the constant nucleons background, Eq.\,\eqref{eq:non_deriv_axion} becomes the mass correction to the axion. The Lagrangian for the axion is
\begin{align}
\label{eq:axion_L}
    \mathcal L = \frac{1}{2}(\partial a)^2 - \frac{1}{2}\left(m_a^2 - \frac{\delta\rho}{4 f_a^2}\right) a^2,
\end{align}
where
\begin{align}
    \delta \rho \equiv \sum_{N = p, n} \delta m_N n_N
\end{align}
and $n_N$ is the number density of the nucleon $N$.
We assume that the nucleon background is sparse, so that $\frac{\delta\rho}{4 f_a^2} \ll m_a^2$.

Using the mass correction, we can estimate the change of momentum of an axion. Assuming that the axion is non-relativistic but its momentum $|\vec{k}|$ is much larger than the mass correction. $n$ is given as
\begin{align}
    n - 1 \simeq \frac{\delta\rho}{8k^2f_a^2} = \frac{\delta\rho}{2 v^2 f_\pi^2 m_\pi^2} \simeq 1.2 \times 10^{-10} \left(\frac{\delta\rho}{0.01\,\text{g}/\text{cm}^3}\right) \left(\frac{v}{10^{-3}}\right)^{-2}
\end{align}
where $v$ is the velocity of the axion. Note that the refractive index is independent of the mass of the axion.

\section{Born approximation}
\label{app:BornApp}
In this section, we review the Born approximation and its validity condition\,\cite{Sakurai:2011zz}.
The scattering problem in quantum mechanics is to solve the Schr\"odinger equation with an appropriate boundary condition, which fixes the wavefunction at infinity as the sum of the plane wave and the outgoing spherical wave. Suppose a Schr\"odinger equation
\begin{align}
    (H_0 + V)|\psi\rangle = E |\psi\rangle,
\end{align}
where $E \equiv k^2 / 2 m$, $m$ is the mass of the particle, $k$ is the momentum of the initial incoming wave, $H_0 \equiv p^2 / 2m$, $p$ is a momentum operator and $V$ is the potential. By solving this Schr\"odinger equation with the boundary condition, we can discuss the scattering process by the potential $V$.

The equation with the boundary condition is rewritten as
\begin{align}
    |\psi\rangle = |k\rangle + (E - H_0 + i\varepsilon)^{-1}V|\psi\rangle,
\end{align}
where $|k\rangle$ is a plane wave state with momentum $k$. 
This is the Lippmann-Schwinger equation\,\cite{Lippmann:1950zz}. The solution of the Lippmann-Schwinger equation can be formally expanded in a power series;
\begin{align}
    |\psi\rangle = &|k\rangle + (E - H_0 + i\varepsilon)^{-1}V|k\rangle \nonumber \\
     &+(E - H_0 + i\varepsilon)^{-1}V(E - H_0 + i\varepsilon)^{-1}V|k\rangle \cdots.
\end{align}
This is called the Born series and the (1st order) Born approximation is an approximation to terminate the series at the first term in $V$;
\begin{align}
    |\psi\rangle \simeq |k\rangle + (E - H_0 + i\varepsilon)^{-1}V|k\rangle.
\end{align}
The physical meaning of the approximation is now clear; the interaction is weak enough for the scattering to occur at most once in the potential and the wavefunction is the superposition of the incoming plane wave and the outgoing scattered wave.

For the Born approximation to be valid, the second term must be much smaller than the first one:
\begin{align}
\label{eq:BornCond}
    \left|\langle x|k\rangle\right| \gg \left|\langle x |(E - H_0 + i\varepsilon)^{-1}V|k\rangle\right|.
\end{align}
The right-hand side can be written as
\begin{align}
    \langle x |(E - &H_0 + i\varepsilon)^{-1}V|k\rangle = \nonumber \\
    &-\frac{2 m}{(2\pi)^\frac32} \int d^3 x' \frac{e^{i k |x - x'| + i\vec{k}\cdot\vec{x}'}}{4\pi |x - x'|} V(x').
\end{align}
As a scatterer, let us think of a spherical potential with radius $R$;
\begin{align}
    V(x) = V_0 \Theta(R - r).
\end{align}
The condition for the Born approximation is that Eq.\,\eqref{eq:BornCond} holds at the origin. It reads as
\begin{align}
    \frac{mR}{k}V_0 \ll 1 \ \ & \text{for}\ \ kR \gg 1 \\
    m R^2 V_0 \ll 1 \ \ & \text{for}\ \ kR \ll 1.
\end{align}
Even if the magnitude of the potential is much weaker than the momentum, the Born approximation is not valid for the large enough sphere.

Let us estimate the cross section for such large but weak potential spheres. Suppose we increase $V_0$ gradually from $0$. For both large or small $k$, when we formally apply the Born approximation, the cross section approaches to the geometrical cross section, $\sigma \sim \pi R^2$ as the inequalities become invalid if the partial waves are appropriately summed over. On the other hand, as an extreme limit, if we take $V_0 \to \infty$, we may exactly solve the Schr\"odinger equation by expanding the scattered wave in spherical waves\,\cite{Sakurai:2011zz}. Then, the cross section is again the geometrical cross section. Thus, it is natural to assume that the cross section is of order $\pi R^2$ even in the intermediate regions.

\bibliography{papers}

\end{document}